\documentclass[12pt]{article}
\usepackage{graphicx}

\newcommand\pubnumber{JLAB-PHY-15-2153}
\newcommand\pubdate{\today}

\def\jlab{Thomas Jefferson National Accelerator Laboratory\\
12000 Jefferson Ave, Newport News, VA 23606 USA}
\def\support{\footnote{Work supported by the U.S. DOE 
under U.S. DOE contract DE-AC05-060R23177}}

\def\Title#1{\begin{center} {\Large #1 } \end{center}}
\def\Author#1{\begin{center}{ \sc #1} \end{center}}
\def\Address#1{\begin{center}{ \it #1} \end{center}}

\newcommand\pubblock{\rightline{\begin{tabular}{l} \pubnumber\\
         \pubdate  \end{tabular}}}
\newenvironment{Abstract}{\begin{quotation}  }{\end{quotation}}
\newenvironment{Presented}{\begin{quotation} \begin{center} 
             PRESENTED AT\end{center}\bigskip 
      \begin{center}\begin{large}}{\end{large}\end{center} \end{quotation}}
\def\Acknowledgements{\bigskip  \bigskip \begin{center} \begin{large}
             \bf ACKNOWLEDGEMENTS \end{large}\end{center}}

\def\beq{\begin{equation}}
\def\eeq#1{\label{#1}\end{equation}}
\def\eeqn{\end{equation}}

\def\beqa{\begin{eqnarray}}
\def\eeqa#1{\label{#1}\end{eqnarray}}
\def\eeqan{\end{eqnarray}}

\let\bar=\overbar

\def\Dslash{\not{\hbox{\kern-4pt $D$}}}
\def\dslash{\not{\hbox{\kern-2pt $\del$}}}

\def\msb{{\bar{\ssstyle M \kern -1pt S}}}

\begin{document}
\begin{titlepage}
\pubblock

\vfill
\Title{Electroweak Measurements of Neutron Densities in PREX and  CREX at JLab, USA}
\vfill
\Author{ Robert Michaels\support}
\Address{\jlab}
\vfill
\begin{Abstract}
Measurement of the parity-violating electron scattering
asymmetry from ${}^{208}$Pb has demonstrated a
new opportunity at Jefferson Lab to measure the 
weak charge distribution and hence pin
down the neutron radius in nuclei in a relatively clean and
model-independent way.  This is because the Z boson of the weak
interaction couples primarily to neutrons.  We will describe the
PREX and CREX experiments on ${}^{208}$Pb and ${}^{48}$Ca respectively.
PREX-I ran in 2010, and CREX and a second run of PREX are currently
in preparation.
\end{Abstract}
\vfill
\begin{Presented}
Twelfth Conference on the Intersections of Particle and Nuclear Physics\\
Vail, Colorado, USA May 19--24, 2015
\end{Presented}
\vfill
\end{titlepage}
\def\thefootnote{\fnsymbol{footnote}}
\setcounter{footnote}{0}

\section{Parity-Violating Measurements of Neutron Densities}
Historically, electromagnetic scattering has accurately 
measured the charge distribution
of nuclei \cite{chargeden,deVries},
providing a detailed picture of the atomic nucleus.
Proton radii have been determined accurately for many nuclei 
using electron scattering experiments \cite{chargeden,deVries,Angeli2013}. 
This accuracy reflects the accuracy of perturbative treatments of the 
electromagnetic process.  The neutron density distribution is more 
difficult to measure accurately because it interacts mainly with 
hadronic probes (pions \cite{pions1}, protons \cite{protons1,protons2,protons3}, 
antiprotons \cite{antiprotons1,antiprotons2}, and 
alphas \cite{ref:alpha1,ref:alpha2}) through nonperturbative interactions, 
the theoretical description of which is model-dependent.
Other approaches to inferring $R_n$ include inelastic scattering excitation of 
giant dipole resonances \cite{ref:GDR,PiekEpja}
and atomic mass fits \cite{Warda2009,Danielewicz2003}.
Neutron radii can also be measured with neutrino-nucleus 
elastic scattering \cite{1207.0693,PRD68.023005}. 

Parity violating electron scattering measures the asymmetry

\begin{equation}
A_{PV} = \frac{\sigma_R - \sigma_L}{
\sigma_R + \sigma_L}
\label{eq:asy}
\end{equation}
where $\sigma_{R(L)}$ is the cross section for right (left)-handed helicity of the 
incident electrons, is very small, of order one part per million (ppm).  

This asymmetry provides a model independent probe of 
neutron densities that is free from most strong interaction uncertainties.  
The $Z^0$ boson, which carries the weak force, couples primarily to neutrons.
In the Born approximation, $A_{PV}$, is
\begin{equation}
A_{PV}\approx \frac{G_FQ^2}{4\pi\alpha\sqrt{2}}
\frac{F_W(Q^2)}{F_{ch}(Q^2)}
\label{eq:born_asy_weakFF}
\end{equation}
where  $G_F$ is the Fermi constant, $\alpha$ the
fine structure constant, and $F_{ch}(Q^2)$ is the Fourier 
transform of the known charge density.  
The asymmetry is proportional to the weak form factor $F_W(Q^2)$.  
This is closely related to the Fourier transform of the neutron density, 
and therefore the neutron density can be extracted from an electro-weak measurement \cite{dds}

\begin{equation}A_{pv}\approx\frac{G_FQ^2}{4\pi\alpha\sqrt{2}}
\Biggl[1-4\sin^2\theta_W-\frac{F_n(Q^2)}{F_{ch}(Q^2)}\Biggr]
\label{eq:born_asy_neutronFF}
\end{equation}

Corrections to the Born approximation from Coulomb distortion 
effects must be included and have been accurately calculated \cite{couldist},  
and other theoretical interpretation issues have been considered in \cite{bigprex}.

The weak form factor is the Fourier transform of the weak charge density $\rho_W(r)$,
\begin{equation}  
F_W(Q^2)=\frac{1}{Q_W}\int d^3r \frac{\sin Q r}{Q r} \rho_W(r),
\label{F(q)}
\end{equation}
and is normalized $F(Q=0)=1$.  The total weak charge of the nucleus is $Q_W=\int d^3r \rho_W(r)$.

Measuring $A_{PV}$ determines the weak form 
factor $F_W(Q^2)$ and from this the neutron radius $R_n$ \cite{ref:prexFF}.
The neutron skin thickness $R_n-R_p$ then follows, 
since $R_p$ is known. 
Finally, the neutron skin thickness constrains the 
density dependence of the symmetry energy 
\cite{rocamaza1307,Vinas1308,EpjaTopical}

Recently, the Lead Radius Experiment (PREX) at Jefferson Laboratory has 
pioneered parity violating measurements of neutron radii and demonstrated 
excellent control of systematic errors \cite{ref:prexI}.  
The experimental configuration for PREX is similar to that used previously 
for studies of the weak form factor of the proton and $^4$He \cite{ref:happex}.  
The Thomas Jefferson National Accelerator Facility provided excellent 
beam quality, while the large spectrometers in Hall A allowed PREX to 
separate elastically and inelastically scattered electrons and 
to greatly reduce backgrounds.           

In this contribution we discuss the PREX-I result \cite{ref:prexI} and the
follow-on measurement PREX-II \cite{ref:prexII} 
and the CREX proposal for $^{48}$Ca \cite{ref:CREX}.
These experiments PREX-II and CREX should measure 
neutron skins with high accuracy.

\section{Experimental Method}
\label{experiment}

The experiments run at Jefferson Lab using the high-resolution
spectrometers (HRS)~\cite{Alcorn:2004sb}
in Hall A, comprising a pair of 3.7 msr spectrometer 
systems with $10^{-4}$ momentum resolution,
which focus elastically scattered electrons onto 
total-absorption detectors in their focal planes. 

A polarized electron beam scatters from a target foil,
and ratios of detected flux to beam current
integrated in the helicity period are formed (so-called ``flux integration"),
and the parity--violating asymmetry in these
ratios computed from the helicity--correlated
difference divided by the sum (eq ~\ref{eq:asy}).
Separate studies at lower rates are required to
measure backgrounds, acceptance, and $Q^2$.  
Polarization is measured once a day by a
M{\o}ller polarimeter, and monitored continuously
with the Compton polarimeter.

The asymmetry is small, of the order of
one or two parts per million (ppm) for the kinematics of interest
for the two nuclei under primary consideration namely, $^{208}$Pb (PREX) and $^{48}$Ca (CREX). 
To have significant impact on our knowledge of skin thicknesses, $A_{PV}$
must be measured with a precision in the range of 3\% or better 
(see fig \ref{fig:R}).
Experiments of this
nature are optimized to the challenges of precision measurement 
of very small asymmetries,
which require high count rates and low noise to achieve
statistical precision as well as a
careful regard for potential systematic errors
associated with helicity
reversal, which must be maintained below the $10^{-8}$ level.

One common feature of all measurements of parity-violation in 
electron scattering is a rapid
flipping of the electron beam helicity, allowing a differential
measurement between opposing
polarization states on a short timescale. The enabling technology 
for these measurements lies
in the semiconductor photo-emission polarized electron source, 
which allows rapid reversal of
the electron polarization while providing high luminosity, 
high polarization, and a high degree
of uniformity between the two beam helicity states.
Developments with the polarized source at
Jefferson Lab are critical to the success 
of this program~\cite{Sinclair:2007ez}.

\par

In a parity experiment,
the asymmetry generally increases with $Q^2$ while the cross
section decreases, which leads to an optimum choice of kinematics.
For parity-violating neutron density experiments, 
the optimum kinematics 
is the point which effectively
minimizes the error in the neutron radius $R_n$.  
This is equivalent to maximizing 
the following product, which is the figure-of-merit (FOM)   

\begin{equation}
FOM =   R \times A^2 \times {\epsilon}^2
\label{eq:FOM}
\end{equation}

Here, $R$ is the scattering rate,
$A$ is the asymmetry, $\epsilon = \frac{dA/A}{dR_n/R_n}$ is the 
the sensitivity of the asymmetry for a small change in $R_n$, 
$dR_n/R_n$ is a fractional change in $R_n$ and $dA/A$ is a 
corresponding fractional change in $A$.  Note that the FOM
defined for many types of parity-violation experiments is $R \times A^2$,
but the neutron-density measurements must also fold in the
sensitivity $\epsilon$.

Given practical constraints on the solid angle of the
HRS, the optimization algorithm favors smaller scattering angles.
Using septum magnets we reach $\sim 5^{\circ}$ scattering angle.
Once the angle is fixed, the optimum energy for elastic scattering can be specified.
Simulations that are performed to design the experiment include 
the Coulomb distortions, as well as radiative losses, 
multiple scattering, and ionization losses in materials, 
together with a model for the tracking of particle 
trajectories through the HRS and septum magnets.

\par The two nuclei of interest for 1\%, or better, $R_n$ measurements 
($^{48}$Ca and $^{208}$Pb) are equally accessible experimentally and have been
very well studied \cite{chargeden,wise,cavendon,emrich,quint}.
These are doubly-magic and have a simple nuclear structure, making
them good candidates for extracting the symmetry energy.
Each nucleus has the advantage that it has a large 
splitting to the first excited state (2.60 MeV for $^{208}$Pb
and 3.84 MeV for $^{48}$Ca), thus lending themselves 
well to the use of a flux integration technique.  

\begin{figure}
\resizebox{0.48\textwidth}{!}{%
\includegraphics{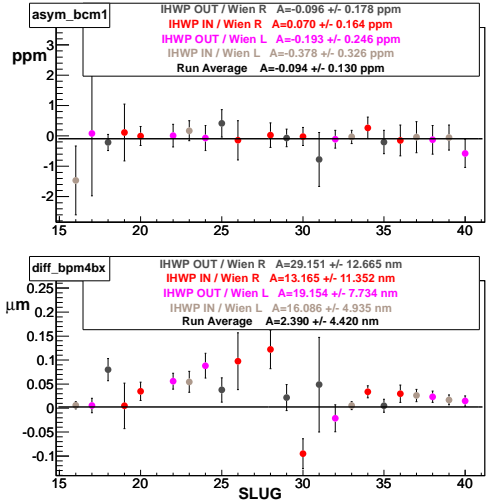}
}
\caption{
PREX-I helicity-correlated charge asymmetries (top) and position 
differences (bottom) on a representative monitor versus slug 
(a slug is $\sim 1$ day of running).  The different colors correspond
to four different combinations of insertable halfwave plate (IHWP) 
and Wien used for slow sign reversal, as 
explained in the text.  
To illustrate the systematics, the data points are plotted without sign correction for the helicity flip.  
The final average with 
all sign corrections is shown by the black horizontal bar and was controlled at 
the 5 nm level averaged over the PREX-I run.  The charge asymmetry was forced 
to zero by the standard feedback system.}
\label{beam_corr}
\end{figure}

To achieve the $10^{-8}$ statistical precision and systematic control 
for $A_{PV}$ measurements requires a precise
control and evaluation of systematic errors, as
has been developed at Jefferson Lab~\cite{ref:happex} 
and elsewhere~\cite{Humensky:2002uv}.
PREX-I was able to achieve overall
asymmetry corrections due to helicity-correlated beam position fluctuations of 
about $40~\mathrm{ppb}$ with position differences $<5~\mathrm{nm}$.  The position/asymmetry
correlations are measured using two independent methods: 
first, directly observing the asymmetry correlations by the natural beam motion and second, by
systematically perturbing the beam through a set of magnetic coils (dithering).
Achieving these small values for the differences was possible in 
part by periodically inserting the half-wave
plate in the injector and flipping the helicity of the beam using a double-Wien
filter which helps them cancel over time.
Fig ~\ref{beam_corr} shows the helicity-correlated charge asymmetries and 
position differences versus time during PREX-I.  A beam current monitor
(BCM) and one representative 
beam position monitor (BPM) is shown; the other BPMs
look similar. Feedback on the charge asymmetry forced it to be zero
within 0.13 ppm.  The utility of the
slow reversals is demonstrated in the BPM difference plot; without them, the 
position differences remained at the $\sim$ 50 nm level (the points without
sign correction) averaged over the experiment; with the reversals,
the differences averaged to the $\sim 5$ nm level (the black lines)
and became a negligible 
correction \cite{KiadThesis,RupeshThesis,LuisThesis,MindyThesis,ZafarThesis}.

\section{PREX-I Result and PREX-II Motivation}
\label{sec.prex}

PREX-I ran in 2010 
and demonstrated successful control of systematic errors,
overcoming many technical challenges, but 
encountered significant loss of beam time due to 
difficulties with vacuum degradation of the target region due
to the high radiation environment \cite{ref:prexI}.
PREX-II is an approved experiment for a followup measurement
with anticipated improvements to take data at a rate
equivalent to the original proposal estimates \cite{ref:prexII}.
PREX measures the parity-violating asymmetry $A_{PV}$ for 
1.06~GeV electrons scattered by about five degrees from $^{208}$Pb.
A major achievement of PREX-I, despite downtimes mentioned above, was
control of the systematic error in $A_{PV}$ at the 2\% level.

The result from PREX-I was \cite{ref:prexI}
\begin{equation}
A_{PV}=0.656 \pm 0.060 ({\rm stat}) \pm 0.014 ({\rm syst})\ {\rm ppm}\, .
\label{A.PREXI}
\end{equation}

\begin{figure}
\resizebox{0.52\textwidth}{!}{%
\includegraphics{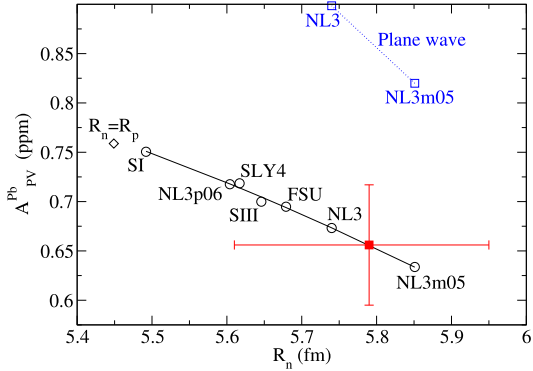}
}
\caption{
Result of the PREX-I experiment (red square) vs neutron point 
radius $R_n$ in ${}^{208}$Pb.  
Distorted-wave calculations for 
seven mean-field neutron densities are 
circles while the diamond marks the expectation for 
$R_n=R_p$ ~\cite{Ban:2010wx}.
\hskip 0.05in References: NL3m05, NL3, and NL3p06 from \cite{ref:nl3},    
FSU from \cite{ref:fsu}, SIII from \cite{ref:siii},
SLY4 from \cite{ref:sly4}, SI from \cite{ref:si}.
The blue squares show plane wave impulse approximation results.}
\label{fig:R}
\end{figure}

This result is displayed in Figure~\ref{fig:R}, in which models predicting the 
point-neutron radius illustrate the correlation of $A_{\rm PV}^{Pb}$ 
and $R_n$~\cite{Ban:2010wx}. 
For this figure, seven non-relativistic and relativistic mean field models 
\cite{ref:nl3,ref:fsu,ref:siii,ref:sly4,ref:si}
were chosen that have charge densities and binding energies in good agreement 
with experiment, and that span a large range in $R_n$.  The weak charge density $\rho_w$ was calculated from model point proton 
$\rho_p$ and neutron $\rho_n$ densities, $\rho_w(r)=q_p\rho_{ch}(r)+q_n\int d^3r'[G_E^p\rho_n+G_E^n\rho_p]$,
using proton $q_p=0.0721$ and neutron $q_n=-0.9878$ weak charges that include radiative corrections.  
Here $G_E^p$ ($G_E^n$) is the Fourier transform of the proton (neutron) electric form factor.  The Dirac equation 
was solved \cite{couldist} for an electron scattering from $\rho_w$ and the experimental $\rho_{ch}$ \cite{chargeden}, 
and the resulting $A_{PV}(\theta)$ integrated over the acceptance to yield the open circles in Fig. \ref{fig:R}.
The importance of Coulomb distortions is emphasized by indicating results from plane-wave calculations, which are not 
all contained within the vertical axis range of the figure.

\section{CREX Proposal}
\label{CREX}

The $^{48}$Ca Radius EXperiment (CREX) was recently approved by the
program advisory committee at Jefferson Lab \cite{ref:CREX}.
The experiment plans to measure the parity-violating asymmetry for elastic
scattering from $^{48}$Ca at E = 2.2~GeV and $\theta=4^{\circ}$.
This will provide a measurement of the weak
charge distribution and hence the neutron
density at one value of Q$^2$ = 0.022 (GeV/c)$^2$. 
It will provide an accuracy in the ${}^{48}$Ca neutron radius $R_n^{48}$
equivalent to $\pm$0.02~fm ($\sim0.6$\%).  
A measurement this precise 
will have a significant impact
on nuclear theory, providing unique experimental input 
to help bridge ab-initio theoretical approaches 
(based on nucleon-nucleon and  three-nucleon forces) and 
the nuclear density functional theory (based on  energy density functionals)
\cite{ref:CREX,ref:HagenNature}
Together with the PREX measurement of $R_n^{208}$, CREX ($R_n^{48}$) 
will provide unique input in such diverse areas such as neutron star structure, 
heavy ion collisions, and atomic parity violation.
A precise measurement on a small nucleus is favorable because it can be
measured at high momentum transfer where the asymmetry is larger 
(for the proposed kinematics, about 2~ppm).
Also, since $^{48}$Ca is neutron-rich it has a relatively large weak charge
and greater sensitivity to $R_n$.  

The significant new apparatus elements for CREX
are the 1 gm/${\rm cm}^2$ ${}^{48}$Ca target and a new $4^{\circ}$ septum magnet. 
The rest of the apparatus is standard equipment and
the methods of section ~\ref{experiment} are applied.
The experiment is designed for $150~\mathrm{\mu A}$
and a 2.2 GeV beam energy, which is a natural beam 
energy at Jefferson Lab (1-pass through the accelerator).
At this energy, the figure-of-merit, which is the total error
in $R_n$ including systematic error, 
optimizes at a scattering angle of $4^{\circ}$.
Table~\ref{table:Experiments} highlights the experimental 
configuration and goals of PREX and CREX.

\begin{table}\centering
\caption{Parameters of the PREX (I and II) and CREX experiments.  }\begin{tabular}{lccc}
 & PREX & CREX \\ \hline \hline
Energy & 1.0 GeV & 2.2 GeV \\
Angle  & 5 degrees & 4 degrees \\
$A_{PV}$  & 0.6 ppm & 2 ppm \\
$1^{\rm st}$ Ex. State & 2.60 MeV & 3.84 MeV \\
beam current & 70 $\mu$A & 150 $\mu$A \\
rate &  1 GHz & 100 MHz \\
run time & 35 days & 45 days \\
$A_{PV}$ precision  & 9\% (PREX-I) 3\% (PREX-II) & 2.4\% \\ 
Error in $R_N$ &  0.06 fm (PREX-II) & 0.02 fm \\ \hline
\end{tabular}
\label{table:Experiments}
\end{table}

\section{Conclusion}

In this contribution we discussed the future measurements 
PREX-II and CREX at Jefferson Lab.  
The parity-violating electron
scattering asymmetry from $^{208}$Pb and $^{48}$Ca 
provide a clean measurement at one $Q^2$ of the 
weak charge of these nuclei and 
are sensitive to the nuclear symmetry energy.
The experiments leverage the advantages Jefferson Lab,
with it's highly stable and precisely controlled electron
beam and the high resolution spectrometers, which are 
uniquely suited to perform these experiments.
Within the next few years, these $R_n$ measurements 
on $^{208}$Pb and $^{48}$Ca 
will provide powerful experimental inputs to tune nuclear models 
of increasing sophistication. 

PREX-I achieved the first electroweak observation,
at the 1.8$\sigma$ level, of the neutron skin of
$^{208}$Pb and successfully demonstrated this technique for
measuring neutron densities, with an excellent control of systematic errors.
The future PREX-II run will reduce the uncertainty by a factor of three,
to $\pm 0.06$ fm in $R_n$.
While PREX-II will put a constraint on the density 
dependence of the symmetry energy (the parameter $L$), 
models predicting neutron radii of medium mass and light nuclei 
are affected by nuclear dynamics beyond $L$. 
CREX will provide new and unique input into the isovector sector
of nuclear theories, and the high precision measurement 
of $R_n$ ($\pm 0.02$ fm) in a doubly-magic nucleus 
with 48 nucleons will help build a critical bridge between 
ab-initio approaches and nuclear DFT.
CREX results can be directly compared to new coupled 
cluster calculations sensitive to three neutron forces \cite{ref:HagenNature}.

\Acknowledgements
The author gratefully acknowledges all the collaborators on
the PREX-II ~\cite{ref:prexII} and CREX ~\cite{ref:CREX} 
proposals and the participants at the
CREX 2013 workshop \cite{CREXworkshop}.

\end{document}